# ORAN Drives Higher Returns on Investments in Urban and Suburban Regions

Priyanka Sharma[a], Edward J. Oughton[b], Aleksan Shanoyan[c]

[a, c] Kansas State University Manhattan KS USA  [b] George Mason University Fairfax VA USA

**Abstract**

This paper provides the first incentive analysis of open radio access networks (ORAN) using game theory. We assess strategic interactions between telecom supply chain stakeholders—mobile network operators (MNOs), network infrastructure suppliers (NIS), and original equipment manufacturers (OEMs)—across three procurement scenarios: (i) Traditional, (ii) Predatory as monolithic radio access networks (MRAN), and (iii) DirectOEM as ORAN. We use random forest, and gradient boosting models to evaluate the optimal margins across urban, suburban, and rural U.S. regions. Results suggest that ORAN deployment consistently demonstrates higher net present value (NPV) of profits in urban and suburban regions, outperforming traditional procurement strategy by 11% to 31%. However, rural areas present lower NPVs across all scenarios, with significant variability at the county level. This analysis offers actionable insights for telecom investment strategies, bridging technical innovation with economic outcomes and addressing strategic supply chain dynamics through a game-theoretic lens.

## 1. Introduction

The global telecommunications mobile industry, once marked by unprecedented growth in demand, is now witnessing stability. The high costs of maintaining and upgrading infrastructure, due to the capital-intensive nature of the industry, have driven increasing consolidation among mobile network operators (MNOs) and network infrastructure providers (NIS) (Genakos, Valletti, and Verboven 2018; Mazzeo 2002; Motta and Tarantino 2021). As of 2023, 95% of the world's population has close proximity to mobile broadband signals, with 4G networks covering 89% of the global population (ITU 2023). However, 2.6 billion people worldwide still remain offline (ITU 2023). A recent study estimates that connecting the unconnected globally would require an investment of approximately $418 billion (Oughton, Amaglobeli, and Moszoro

2023). At the same time, the accelerating adoption of Internet of Things (IoT) technologies is creating new opportunities in wireless connectivity (Ericsson 2024). IoT applications, particularly in sectors like logistics, agriculture, and smart cities, are reshaping data requirements and fostering innovation within the telecommunications industry. In the United States, wireless networks supported approximately 78 petabytes of data in 2023, an 89% increase since 2021 by the cellular telecommunications and internet association (CTIA 2024). To meet this demand, U.S. wireless carriers, also known as MNOs, have invested a cumulative $705 billion to date, including $30 billion in 2023 alone (CTIA 2024).

Growing demand for data from business specific network solutions has placed pressure on the network infrastructure industry. To meet the demand for these new use cases and to retain the existing market share, MNOs must provide more innovative and business-specific solutions. It necessitate significant upgrades in network capacity and performance, requiring firms to physically upgrade their networks and rethink their broader network strategies (Rendon Schneir et al. 2019). Generally, network upgrade costs for MNOs are largely driven by the radio access network (RAN), which accounts for up to 70% of the total network cost (GSMA 2019).

Traditional RAN infrastructure approaches, commonly known as monolithic radio access networks (MRAN), struggle to meet the growing demand for virtualized, innovative and business specific network solutions. Because MRAN relies on an oligopolistic market structure where a handful of NIS, such as Nokia, Ericsson, Huwaei, ZTE., offer custom hardware bundled with proprietary software. The business model of NIS companies, in essence, is to design equipment, buy components, and manufacture final units from OEMs. This market structure restricts competition and limits MNOs' negotiation power for new service deployments and upgrades, often leading to expensive, inflexible generational upgrades with reduced innovation due to vendor lock-in. This is leading to ongoing technical research developments that aim to develop virtualized, software-defined networks with intelligent machine-controlled data traffic management (Shu-ping Yeh et al. 2024).

On the policy side, the dominance of a few global suppliers in MRAN exposes MNOs to significant supply chain vulnerabilities. Therefore, the demand for indigenous and open telecom infrastructure supply chains is increasing (Boyens et al. 2021; Department for Science, Innovation and Technology 2024). To fulfill such requirements, a growing body of stakeholders are seeking domestic, multi-vendor open ecosystems of suppliers. Its aim is to enhance national security and foster domestic innovation, creating high-skilled jobs and reducing dependency on imported technologies (Kim, Eom, and Lee 2023; NTIA 2024a, 2024b). Therefore, many government agencies introduced policies that foster and develop supply chain diversity for RAN upgrades. For example, under the Secure and Trusted Communications Networks Act of 2019, the United States Federal Communications Commission (FCC) promotes open supply chain options (FCC, Federal Communications Commission 2022). In Japan, the ministry of internal affairs and communications (MIC) promotes the adoption of ORAN through its Beyond 5G Promotion Strategy (B5G 2020). EU policies for 5G and 6G also take a 'technology-specific' approach, focusing on demand-side innovation alongside supply-side efforts(Rossi 2024).

In response, many industry groups, MNOs, neutral hosts, integrators, businesses, and non-government groups collaborated to design an open infrastructure RAN supply chain (Oughton et al. 2024). Various industry alliances, such as the O-RAN alliance (GSMA 2020), Meta's Telecom Infra Project (TIP) (Meta 2016), small cell forum (Femto Forum 2007) have designed open interface standards and the guidelines for the disaggregation of software from hardware also called as open radio access network (ORAN).  Figure 1 shows the detailed comparison of architecture and supply chain. Various technical studies suggest that ORAN increases flexibility and fosters innovation, thanks to the ability of ORAN software to be interfaced with commercial off-the-shelf (COTS) hardware, facilitating over-the-air (OTA) virtual updates (Bonati et al. 2020; Polese et al. 2023b).

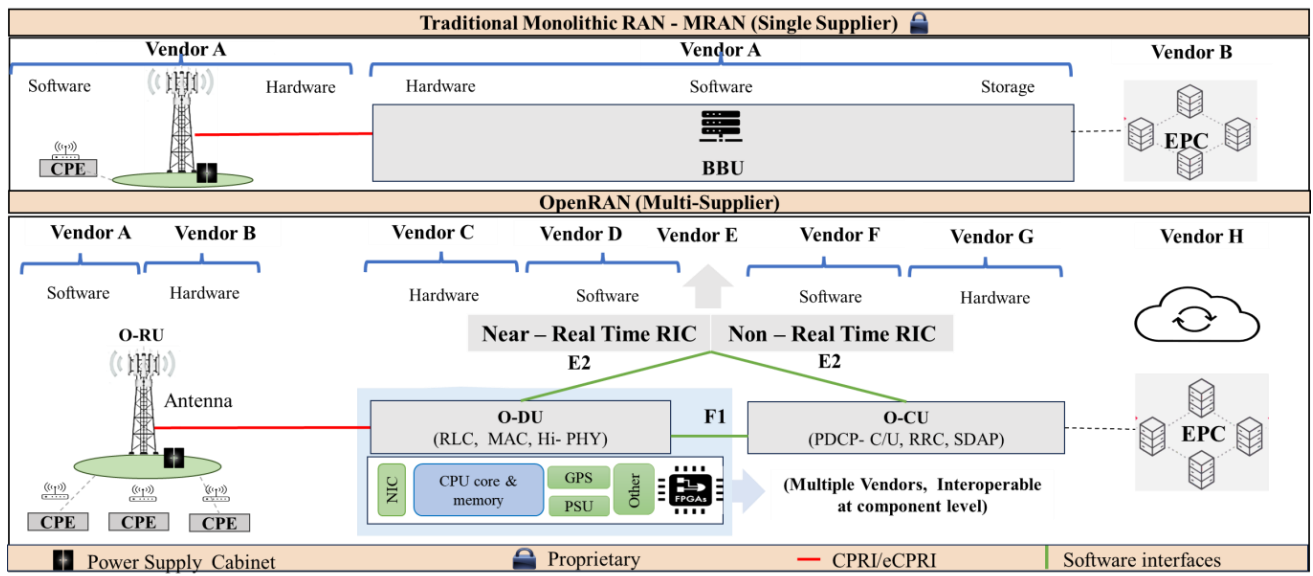

Figure 1 - MRAN vs ORAN Supply Chain by Network Architecture

Thus, ORAN represents a paradigm shift towards disaggregated, interoperable, and vendor-neutral RAN architectures. This shift offers both technical and business decentralization. On the technical side, network function virtualization (NFV) and software-defined networking (SDN) enable dynamic allocation of network resources, reducing the need for large upfront investments and allowing more agile responses to fluctuating demand (Hiba Hojeij et al. 2023). In terms of supply chain,

ORAN fosters a diverse ecosystem of vendors and solutions, addressing national security concerns and promoting indigenous technological development. Recent innovations further enhance ORAN's potential. HexRAN focuses on fully customizable base station system design (A. Kak et al. 2023), while multi-vendor, programmable testbeds utilizing NVIDIA ARC and OpenAirInterface facilitate agile network designs and scalable solutions (Villa et al. 2023). Moreover, deployment of ORAN technologies in rural and underserved regions promises in bridging the digital divide (F. Slyne et al. 2024). It allows for targeted bandwidth and latency implementations, making it especially valuable for specialized business operations and connectivity initiatives. These sector-specific drivers highlight ORAN's role not just as a technical evolution, but as an enabler of digital transformation across diverse industries.

However, ORAN do introduce some technical challenges beyond market structure complexities (Cambini and Jiang 2009). First, technical issues may arise in multi-vendor environments, leading to conflicts in traffic steering and resource allocation for which conflict management systems are being developed (Abdul Wadud, Fatemeh Golpayegani, and Nima Afraz 2023). Second, the multi-sided nature of the Open RAN ecosystem, involving MNOs, vendors, and software suppliers, adds another layer of complexity to identifying and addressing anticompetitive behavior as market dynamics shift, with potential predatory pricing by incumbent suppliers. In any oligopolistic market with few firms, the pricing decision of each firm significantly impacts the market, despite the availability of close substitutes from competitors. Similarly, RAN suppliers as an oligopoly may collude explicitly or implicitly to set prices that maximize their collective profits. This behavior leads to higher prices and lower output than a more competitive market structure. That is why ORAN, which intends to convert the oligopolistic market into a more competitive supplier market, is facing predatory pricing from MRAN suppliers. This dynamic motivates the need for a game theoretic approach to study the strategic interactions between MNOs and suppliers in the ORAN ecosystem. Therefore, a key area of concern in the adoption of ORAN is the possibility that incumbents may engage in Predatory Pricing or other exclusionary strategies, which could impact the total cost of ownership (TCO).

Therefore, in this capital-intensive and rapidly evolving RAN landscape, MNOs must carefully allocate investments to meet growing data demand while maintaining cost efficiency (Naudts et al. 2016). Various studies emphasize that equilibrium prices and potential revenue are heavily influenced by suppliers' positions within the network, underscoring the importance of region-specific investment strategies for MNOs to remain competitive (Toka et al. 2021). However, a significant issue in quantitative decision-making is the lack of available data on costs and prices for each player in the supply chain, often due to closed book contracting terms enforced by non-disclosure agreements (NDAs) in commercial contracts. This lack of transparency exacerbates information asymmetry and moral hazard, reducing incentives for innovation and cost efficiencies across the supply chain.

A significant research gap exists in understanding the strategic incentives required for the adoption of ORAN. Specifically, there is a need for further examination of the dynamics between OEMs that manufacture and supply hardware components, NIS that offer complete network solutions as services, and MNOs. These interactions, particularly in the context of ORAN, should be compared with MRAN under different procurement scenarios to provide more insights into investment strategies. Understanding incentive structures in ORAN is critical for realizing its promised benefits of increased competition, innovation, and cost reduction in telecommunications infrastructure, particularly as demand for low-latency, high-bandwidth services grow in underserved regions. While technical specifications and field trials for ORAN mature, the financial incentives driving the adoption by key stakeholders significantly impact regional deployment. Properly aligned stakeholder incentives can help bridge the rural-urban digital divide to coverage, capacity and latency, while reducing government intervention and making ORAN deployment sustainable.

Given this market situation, this research investigates three key questions:

1. How do market structures (single-supplier MRAN vs. multi-supplier ORAN) influence stakeholder incentives and profit distribution across regions?

2. What strategies are available to each stakeholder in MRAN and ORAN markets, and how do they influence profits ?

3. What procurement strategies optimize stakeholder payoffs in MRAN and ORAN markets, and how do these strategies vary with regiontype ?

To address these questions, Section II of this paper reviews the relevant literature on telecom infrastructure optimization strategies, laying the groundwork for the game-theoretic analysis presented in Section III. This is followed by Section IV, which details the United States case used to evaluate the financial and strategic outcomes of different procurement structures in the U.S. telecom industry, comparing MRAN and ORAN incentives. Finally, Section V briefs the conclusion and discussion.

**2. Literature Review**

Many earlier studies on telecom infrastructure investment strategies primarily focused on traditional, vertically integrated network models heavily reliant on proprietary solutions from a limited number of suppliers ( Whalley, 2002; Fransman,

2003; Jackie Krafft & Krafft, 2003; Li & Okamoto, 2009). Subsequently, this literature review covers a range of topics related to the research questions, providing important background on the current state of knowledge and analysis pertaining to the mobile network equipment market.

There are three ways to enhance a wireless network. Firstly, upgrading the RAN technology to a more spectrally efficient one, such as going from 4G to 5G, resulting in increased bits per hertz per second. Secondly, adding additional spectrum bandwidth to sites to increase overall throughput. Or finally, building more sites to increase the density of assets, increasing the spectral reuse of spectrum resources. In addition to these three strategies, there are also a range of more minor options to improve the efficiency of wireless networks, ranging from enhanced intercell interference coordination (Lopez-Perez et al. 2011), through cloudification and network slicing (Papageorgiou et al. 2020). Equally, the business model utilized by MNOs can also have a substantial impact on the delivery of ultra-dense wireless networks (Chen et al. 2016), including a range of more basic passive and active infrastructure-sharing options to neutral host models (Cave 2018), with estimated cost savings of up to USD 940 million in South Korea as an example. The emergence of 5G has represented a paradigm shift in technology and delivery options for MNOs, significantly reshaping the competitive landscape of the telecom industry (Lehr, Queder, and Haucap 2021), unlike the incremental improvements of previous generations. One example is the shift towards local private 5G networks, either utilizing locally licensed or shared spectrum bands (Matinmikko-Blue et al. 2018).

The approaches articulated can be utilized to address the specific challenges associated with certain deployment environments, from dense urban to rural and remote areas. E.g. for dense urban areas, research has focused on millimeter-wave frequencies, distributed antenna systems (DAS), the joint deployment of small and macro cells (Araujo et al. 2018), small cells (Cano et al. 2019; Nikam et al. 2020), and massive multiple-input multiple-output (MIMO)(Van Rompaey and Moonen 2023; Zhao et al. 2024) to increase capacity and reduce per-bit costs. Meanwhile, rural areas see potential in integrating satellite and high-altitude platform systems for wide area coverage (Osoro et al. 2024; Osoro and Oughton 2021; Yaacoub and Alouini 2020) and the use of Unmanned Aerial Vehicle (UAV)-based architectures to extend coverage (Chiaraviglio et al. 2017). NeutRAN architecture is also studied as a way to enhance rural connectivity by enabling spectrum-sharing and elastic resource allocation (Leonardo Bonati et al. 2023) and low power wide area network (LPWAN) technologies for large-scale IoT deployments for rural distance ranges (Mekki et al. 2019). However, (Oughton et al. 2019) reported that the rural-urban digital divide still exists, finding that rural areas have fewer high-speed fixed and mobile providers but more slower-speed fixed providers than urban areas.

Therefore, alongside network optimization strategies, the demand for modular and customizable network deployments like ORAN gained momentum. Various technical studies demonstrated ORAN's cross-industry potential (Jordi Mongay Batalla et al. 2022). dApps have also been proposed as complements to xApps and rApps to enable real-time control for cases like beam management and user scheduling (Salvatore D'Oro et al., 2022). Federated deep reinforcement learning techniques emerged as effective solutions for automating RAN slicing in decentralized ORAN architectures (Amine Abouaomar et al. 2023). Additionally, the combination of digital twin technology with ORAN is seen as a promising direction for resilient 6G RAN deployment (Antonino Masaracchia et al. 2023). Furthermore, the practical applications of ORAN, such as modular traffic steering in open networking, have been demonstrated (Dryjański, Kułacz, and Kliks 2021). Also, programmable intelligence frameworks for traffic steering further enhance resource allocation within ORAN (Andrea Lacava et al. 2023).

Consequently, many researchers have increasingly focused on studying open ecosystems and modular network solutions (Garcia-Saavedra and Costa-Perez 2021), demonstrating the potential of data-driven optimization in ORAN networks (Bonati et al. 2021). These innovations have shown the capability to significantly reshape the competitive landscape of the telecom industry (Giannoulakis et al. 2016; Paglierani et al. 2020; Greer, Bohn, and Zhang 2022; Wypiór, Klinkowski, and Michalski 2022). Moreover, new developments, such as semantic communications in ORAN systems, have demonstrated potential integration with 6G, enabling intelligent, knowledge-driven network infrastructures (Akyildiz, Kak, and Nie 2020; Peizheng Li and Adnan Aijaz 2023). Research has further highlighted how adopting open models can reduce dependency on proprietary solutions and foster increased market competition (Oughton and Lehr 2021). Expanding on this, a comprehensive review of ORAN systems has added technical depth by covering the complete architecture and design (Rouwet 2022).

Further studies have delved into the architecture, interfaces, and security challenges of ORAN, demonstrating its potential to reshape market dynamics (Oughton et al. 2024; Polese et al. 2023b). Machine learning applications in ORAN, particularly at the network edge, have also been investigated (Polese et al. 2021, 2023a). The deployment of deep reinforcement learning models has been proven pivotal in optimizing ORAN's resource allocation, emphasizing the potential of AI-driven approaches (Fatemeh Lotfi, F. Afghah, and J. Ashdown 2023). Additionally, research into multi-agent team learning in disaggregated, virtualized ORAN environments has added to the understanding of the competitive advantages of open models (Rivera, Mollahasani, and Erol-Kantarci 2020). Moreover, the integration of network digital twins into ORAN environments enhances the ability to simulate and predict network behaviors, improving the overall robustness and adaptability of these systems (Javad Mirzaei, Ibrahim Abualhaol, and Gwenael Poitau 2023). Advancements like ScalO-RAN optimize AI-based ORAN applications to meet stringent latency requirements (Maxenti et al. 2024). while benchmarks such as ORAN-Bench-13K highlight the potential of large language models (LLMs) in ORAN for network analytics, anomaly detection, and code generation (Gajjar and Shah 2024). Furthermore, adaptive streaming technologies like 360-ADAPT leverage ORAN to enhance quality in streaming 360° opera content by prioritizing audio over video for better use of resources and energy efficiency (Simiscuka et al. 2024). Additionally, frameworks like PandORA automate DRL agent design for dynamic control in ORAN, achieving significant network performance improvements (Tsampazi et al. 2024). These studies highlight the growing need for more modular and interoperable alternatives in telecom infrastructure, bringing increased attention to research on open and flexible network architecture. Finally, a comprehensive review of the state-of-the-art in open, programmable, and virtualized 5G networks has outlined the road ahead for further innovations, including AI-driven network orchestration and edge solutions through next-generation open interfaces (Bonati et al. 2020).

In terms of cost- benefit evaluations, several scholars have developed frameworks to assess alternative market and industry structures. These frameworks help evaluate the impact of open ecosystems and technological innovations on traditional market players, providing insights into how market dynamics might evolve under different scenarios (Cano et al. 2019; Paglierani et al. 2020). For instance, a scenario-based assessment of the future supply and demand for mobile telecommunications infrastructure enables a comparative understanding of capacity, coverage, and cost (Oughton et al. 2018; Oughton and Frias 2018). Additionally, various studies explored policy choices to keep 4G and 5G broadband affordable (Oughton et al. 2022), while other studies examined the techno-economic cost implications of 5G (Koratagere Anantha Kumar and Oughton 2023; Kumar and Oughton 2023; Oughton and Frias 2018).

When analyzing market challenges, various foundational frameworks (Areeda and Turner 1975; Ordover and Willig 1981) provide insights into typical Predatory Pricing strategies in such market structures. However, the latest argument on predation in multi-sided markets suggests that market analyses may fail to capture the complexities of multi-sided markets, such as the telecom industry (Jullien and Sand-Zantman 2021). Similarly, another study on platform dynamics suggests that incumbents in network effect-driven industries may rationally engage in below-cost pricing for extended periods to prevent the adoption of new platforms (Cabral 2019). These studies are directly relevant to the impact of predatory pricing in ORAN. Although helpful, these frameworks may not fully capture the complexities of modern telecom markets, which are characterized by strong network effects and rapid technological changes.

Moreover, beyond pricing strategies, recent research has emphasized a new concept, "predatory innovation", where incumbents deter competition through control over ecosystem design and product innovation rather than direct price manipulation. These non-price strategies—such as market tipping, product collusion, and ecosystem control—allow incumbents to maintain dominance (van Oosten and Onderstal 2023). In the context of ORAN, this dominance may be reinforced through long-term exclusive contracts and control over critical network components (Calzolari and Denicolò 2015). This multifaceted approach underscores the need for a more nuanced antitrust analysis, as argued by (Sidak and Teece 2009). The market dynamics of the RAN industry provide strong motivation for utilizing a game theoretic approach, as already demonstrated for network sharing strategies in multi-operator cellular networks (Bousia et al. 2016).

Several studies have employed game-theoretic models in the telecommunications sector. For example, Bertrand game to model spectrum pricing competition among primary service NIS in cognitive radio networks, incorporating quality of service constraints (Niyato and Hossain 2008). Stackelberg game is used to optimize data offloading between LTE and Wi-Fi networks and to find economic incentives for balancing traffic load (Anbalagan et al. 2019). Applied evolutionary game theory and the hoteling model are used to analyze collusion and competition strategies among telecom operators (Wen and Fu 2014). These studies simulate strategic interactions to enhance decision-making in competitive telecommunication industry environments. For the satellite constellation industry, a game-theoretic framework is used to analyze competition among satellite constellation operators in low-Earth orbit (Guyot, Rao, and Rouillon 2023). It is a two-stage game where operators choose constellation designs and then compete on price. This approach revealed how oligopolistic competition can lead to suboptimal outcomes in terms of economic welfare and resource allocation. In our study, we are applying a three-player vertical Stackelberg game to the ORAN ecosystem.

## 3. Methodology

This section introduces a game theory-based framework focusing on the interactions between key supply chain players—MNOs, NIS, and OEMs. Rather than addressing infrastructure sharing between MNOs, we explore the strategic leadership dynamics between these stakeholders. The goal is to develop a decision-making framework to understand how each RAN entity optimizes its strategies based on the actions of others in heterogeneous regions and negotiations.

As the literature suggests, understanding telecom value chain profitability, particularly in the context of ORAN, requires the consideration of both financial and non-financial factors, such as the predatory behavior of existing suppliers or signaling to switch. Also, telecom stakeholders' profits depend on various demographic and technological factors, which vary significantly across different regions, such as urban, suburban, and rural areas. Therefore, it is important to calculate the expected profit function for each stakeholder—MNOs, NIS, and OEMs—to capture individual profits, interdependencies, and strategic interactions within the ecosystem. Each stakeholder plays a critical role in the deployment and success of network infrastructure, with their decisions being closely linked. By calculating the profits for each entity, we aim to understand how one stakeholder's investment incentives in the form of profit impact the outcomes for the others from a game theory perspective.

| Player | Strategies/Action | | |
|---|---|---|---|
| OEM (C) | Continue with MRAN | Invest in manufacturing ORAN compliant components | |
| NIS (B) | Traditional Pricing to MRAN | Predatory Pricing to counter ORAN and Offer MRAN | Transition products to ORAN deployment standards |
| MNO (A) | Continue using MRAN | Demand ORAN but takes Predatory Priced MRAN | Switch to ORAN |
| Scenarios -> | S1: Traditional | S2: Predatory | S3: DirectOEM |

Table 1 - Modelling RAN Transactions Strategies

A key aspect of this game is to shift price dominance from NIS to MNOs under competitive industry structure. Since MNOs are the customers, they should be able to bid—or not bid—during the procurement process. This shift creates three strategic procurement decision options with distinct pricing scenarios, each affecting individual stakeholders' margins as well .

i) The first scenario is a Traditional Pricing scenario ($s_1$). It involves a standard procurement process where suppliers and OEMs offer existing market-rate prices. This approach maintains stable cost structures and margins, fostering long-term supplier relationships (Cambini and Jiang 2009).

ii) Second is the Predatory Pricing scenario ($s_2$). It occurs when MNOs signal their intent to switch suppliers, prompting NIS to adopt aggressive pricing tactics to retain contracts. Although this strategy may result in lower unit prices and reduced supplier margins in the short term, it may negatively impact RAN's long-term performance, potentially leading to decreased research and development (R&D) investments and operational efficiency.

iii) Last is the Direct OEM Pricing scenario ($s_3$). In this scenario, MNOs bypass NIS altogether and directly procure

equipment from OEMs, which is called the ORAN ecosystem. This requires upfront investment and time to get the necessary hardware and software, with added maintenance responsibility.

Therefore, the objective function of our study is to evaluate and compare supply chain profits as incentives by region type under each scenario. For $s_1, s_2,$ objective function can be expressed as a sequential profit maximization under traditional and predatory supply chain:

Player B - NIS (leader)

$$m_{sr}^* = arg \max_{m_{sr} \in [m_{sr}^{min}, m_{sr}^{max}]} \sum_{t=1}^{N} \beta^t \Pi_{sr}(m_{sr})$$

(1a)

Player C - OEM (follower)

$$m_{or}^*(m_{sr}) = arg \max_{m_{or} \in [m_{or}^{min}, m_{or}^{max}]} \sum_{t=1}^{N} \beta^t \Pi_{mr}(m_{or}, m_{sr}^*)$$

Player A (passive) (1b)

$$\Pi_{mr} = Q_{mr}(P_{mr} - C_{nr} - m_{nr} - C_{or} - m_{or})$$

(2c)

Whereas the objective function changed for ORAN scenario i.e. OEMDirect S3 scenario as a joint profit maximization as it's a competitive market now. The MNO bypasses the NIS and directly procures infrastructure components from the OEM. The best response functions for MNO and OEM are as below. The best response functions can be described as the price and site as quantities that maximize the joint industry profits.

$$\text{ORAN}(s_3), \ \max \sum_{t=1}^{N} \beta^t [\Pi_{mr} + \Pi_{or}]$$

(3)

where:

- $s_1$ is a Traditional Pricing scenario.
- $s_2$ is the Predatory Pricing scenario.
- $s_3$ is the Direct OEM Pricing scenario.
- $\Pi_{mr}$ represents the profits of the MNOs in region r.
- $\Pi_{sr}$ represents the profits of the NIS in region r.

- $\Pi_{or}$ represents the profits of the OEMs (Original Equipment Manufacturers) region r.
- t denotes each period or stage of the market dynamics being considered, up to N periods.
- $\beta^t$ is the discount factor with $0 < \beta < 1$.
- $m_{nr*}$, $m_{or*}$ : Margins set by Player NIS and Player OEM

### 3.1. Mobile Network Operators (MNO)

The expected profit $\Pi_{mr}$ for an MNO in the region, r can be expressed as:

$$\Pi_{mr} = R_{mr,t=0} - C_{mr,t=0} + \sum_{t=1}^{N-1} \frac{R_{mrt} - C_{mrt}}{(1+d)^t}$$

( 4 )

where $R_{m,r}$ is the revenue generated for MNO m in region type r and $C_{mr}$ is the cost associated with the investment in region type r, including capital expenditure (Capex), i.e. $C_{mr,t=0}$, operational expenditure (Opex) i.e. $C_{mrt}$ in region type r and d is the discount rate.

**Revenue Function**

The revenue model in both competitive market and oligopolistic market scenarios considers various smartphone adoption and demographic factors, where $ARPU_{rt}$ is the average revenue per user in region type r, $M_{rt}$ represents the market share of MNO in region type r at year t, indicating the proportion of the total market in region type r at year t captured by MNO, and $U_{rt}$ represents the active smartphone users in region type r at year t, directly affecting revenue opportunities due to the potential number of customers which can be modeled as follows:

*Ideal Competitive Market* - In this scenario, MNOs compete independently, and the revenue is calculated considering the individual market share $M_{rt}$ as:

$$R_{mrt} = ARPU_{rt} \times M_{rt} \times U_{rt}$$

( 5a )

*Oligopolistic Existing Market* - In this scenario, MNOs cooperate or collude to divide market regions. As a result, the revenue model reflects the aggregated revenue for all MNOs in region r. In such a market structure, there is no need to consider individual market shares for each MNO type in the region. Hence, the expected profit function for MNO in region r is:

$$R_{mrt} = ARPU_{rt} \times U_{rt}$$

(4b)

a) **Adjusted ARPU$_{rt}$ using Median Household Income** - The average revenue per user (ARPU) in region type r is adjusted by the median household income (MHI$_r$) to account for the influence of household income in region type r on ARPU. It serves as a proxy for the actual revenue per user for each region type. A recent study shows that markets with higher median household income tend to have higher market shares for expensive phone plans (Elliott et al. 2023). This adjustment reflects the spending power of users in different region types on mobile services, i.e., higher-income regions potentially supporting higher ARPU.

$$ARPU_{rt} = \text{national ARPU}_t \times \left(\frac{MHI_{rt}}{NHI_t}\right)$$

(6)

where:

- MH$_{rt}$: Median Household Income in region type r in year t.

- NHI$_t$: National Household Income in year t.

b) **Total Users, U$_{rt}$** - The total number of active smartphone users (Ut,r) in region type r at time t is calculated by considering the initial population, population growth rate, smartphone penetration rate, and active users rate. The formula integrates the initial values and their respective growth rates to reflect changes over time:

$$U_{rt} = \left[P_{r,t=0} \times (1 + g_r)^{t-1}\right] \times S_r \times A_r$$

(7)

where:

- $P_r(0)$ is the initial population in region r at time t = 0.

- $g_r$ is the average population growth rate in region type r.

- $S_r$ is the smartphone penetration rate in region type r.

- $A_r$ is the active user rate in region type r.

**Cost Function**

Total Cost of Ownership (TCO) in the RAN industry is calculated by assessing the NPV of RAN infrastructure investments, where Capex is the fixed cost in year 0, and Opex is modeled as a fixed percentage of Capex used to measure the maintenance of that Capex over the life of the product. Mathematically, the cost function $C_{mrt}$ is TCO for MNO to procure total telecom sites and to maintain them in region r that includes both the initial investment as Capex $C_{mr,t=0}$ and the ongoing operational and maintenance expenses as Opex $C_{mr,t>0}$ The total number of sites in that region is below.

$$C_{mr,t=0} = \text{Sites}_{rt} \times P_s$$

$$C_{mr,t>0} = \text{Sites}_{rt} \times P_s \times \text{OpexRate}$$

(8)

Price $P_s$ of each Site is determined by the procurement strategy and region type that also influences the Opex as it is typically the % of Capex.

a) **Total Sites, Sites$_{rt}$** - The total number of sites required over the year in region r to meet the total data demand is measured by using both sites needed based on data demand, Sites$_{rt,d}$ and the site needed based on coverage by area, Sites$_{rt,a}$, which can be calculated using the following formulae:

$$\text{Sites}_{rt,d} = \frac{D_{rt}}{S.C._{rt}}$$

(9)

In addition to data demand, the total number of sites required is also influenced by the need to cover the geographical area. The coverage area of each site $A_r$, measured in square km, depends on the coverage radius $R_r$.

$$\text{Sites}_{rt,a} = \frac{A_{total}}{A_r}$$

(10)

$$A_r = \pi \times R_r^2 (1-I_r).$$

(11)

To ensure that both the data capacity and coverage requirements are met, the total number of sites required in a region can be adjusted to account for the geographical area. The adjusted formula for the total sites needed in a region r is as follows:

$$\text{Sites}_{rt} = \max(\text{Sites}_{rt,d}, \text{Sites}_{rt,a})$$

(12)

This formula ensures that the number of sites deployed in each region is sufficient to meet both the total data demand and provide adequate coverage for the entire geographical area. In rural regions, where coverage may be the primary constraint, the number of sites is often determined by the coverage requirement. At the same time, data demand typically drives the need for additional sites in urban areas.

Where:

- $D_{rt}$ is the total data demand in region r at year t, measured in bits per year.
- $SC_{rt}$ is the site capacity in region r in year t, measured in bits per year.
- $A_{total}$ is the total land area of the region, measured in square km.
- $R_r$ is the coverage radius of the site in region r, as per the configurations selected in section 0, measured in

square km.

- $A_r$ is the geographical coverage area per site in region r, based on the coverage radius $R_r$, measured in square km.

- $I_r$ Interference factor ranges from 0 (no interference) to 1 (complete interference), which reduces the effective coverage area of a site.

- $Sites_{rt,d}$ is the number of sites required to meet data demand in region r at year t.

- $Sites_{rt,a}$ is the number of sites needed based on the geographical area of the region r at yeat t.

- $Sites_{rt}$ Number of sites required to meet both data demand and geographical coverage constraints.

b) **Total Data Demand, $D_{rt}$** - To estimate the initial investment cost, it is crucial to calculate the total data demand first. The overall demand reflects the annual data needs based on user data consumption patterns and can be expressed as:

$$D_{rt} = U_{rt} \times UD_r^{Users} \times 8$$

(13)

where:

- $U_{rt}$ is the number of active smartphone users in region r at year t.
- $UD_r^{Users}$ is the yearly data demand per user (in bytes/user/) in region r.
- 8 is the conversion factor from bytes to bits.
- $D_{rt}$ is the total data demand in region r at year t, measured in bits per year.

c) **Site Capacity Calculation, $SC_r$** – The capacity of each site is calculated using spectral efficiency, bandwidth, and number of sectors for each site configuration decided based on the region type. It helps in understanding the total data throughput capability of sites, which is crucial in finding the required number of sites in each region. The capacity is instantaneous, measured in seconds for each site, and remains constant for the years of the site's life. For our study, we convert it into bits per year. Therefore, for each region type r, the site capacity is calculated based on the site type used to fulfill the demand as:

$$\text{S.C.}_r = B_r \times SE_r \times NS_r \times 2592000 \times 12 \times 10^6$$

(14)

- $B_r$ is the bandwidth in region r (MHz), $\times\ 10^6$ to convert into Hz.
- $SE_r$ is the spectral efficiency in region r (bps/Hz).
- $NS_r$ is the number of sectors in region r.
- 2592000 is the number of seconds a month (assuming 30 days in a month.)
- 12 is the conversion factor for monthly to yearly data demand.
- $SC_{rt}$ is the site capacity in region r in year t, measured in bits per year.

### 3.2. Network Infrastructure Suppliers (NIS)

The second important stakeholder in the RAN supply chain is NIS, denoted as s, which provides the necessary hardware and software solutions to MNOs. Their expected profit over the N-year period is influenced by the contracts they secure with MNOs, the pricing of their products, and the costs associated with design customization and R&D, as they are not directly producing the products. In an oligopolistic market, one type of region is supplied by one exclusive NIS. Assuming the same margin is applied to capex and opex. The expected profit for a supplier over the N-year period in region r can be expressed as:

$$\Pi_{sr} = \text{Sites}_t \times (P_s - P_o) + \sum_{t=1}^{N-1} \frac{\text{Sites}_t \times (P_s - P_o) \times \text{OpexRate}_{rt}}{(1+d)^t}$$

(15)

where:

- $\text{Sites}_t$ represents the number of infrastructure sites supplied to MNOs in year 0 as capex.
- $P_s - P_o = M_s$ represents the NIS margin (profit per unit) based on the selected strategy in region r.

### 3.3. Original Equipment Manufacturers (OEM)

The third stakeholder category is OEMs, denoted as o, responsible for providing the key components and equipment used in RAN infrastructure along with the baseline software interfaces. Multiple OEMs collaborate with contract manufacturers to supply the finished equipment to NIS in the MRAN structure and provide white box hardware in the ORAN structure.

In MRAN, the software is typically developed in-house by the NIS. In contrast, in ORAN, the software interfaces are provided independently through ORAN alliance groups, with OEMs also providing baseline software. If we consider OEMs and contract manufacturers as one category of stakeholders, their expected profit over the N-year period is based on production efficiency and cost management. Also, revenue is a one-time Capex in year 0. Therefore, expected profit for an OEM for region r can be expressed as:

$$\Pi_{or} = \text{Sites}_t \times M_o$$

(16)

where:

- $\text{Sites}_t$ is the production output (equipment units for sites) of the OEM in year 0 as capex.

- $P_o - P_{cost} = M_o$ represents the OEM margin (profit per unit) based on the selected strategy in region r.

Therefore, The OEM's total profit depends on the production volume and the margin, both influenced by the MNO procurement strategy that shapes the revenue and cost structure. Hence, the choice of strategy directly impacts the revenue and cost structure of the OEM, affecting the overall profitability of OEMs and related next-tier industries.

### 3.4. Stackelberg Competition in RAN Supply Chain Under Predatory Scenario

Various studies have demonstrated that a dominant actor maximizes utility even in decentralized supply chains, and contract compliance can shift local equilibria toward global optimality(Arda and Hennet 2005). Therefore, we utilize a game-theoretic approach to evaluate the decision preferences in such scenarios, using profits as payoffs. Our study aims to model and analyze these scenarios using a Stackelberg game-theoretic approach, building on the economic foundation of existing oligopoly models (Fudenberg and Tirole 2010) & (Bonatti, Cisternas, and Toikka 2016; Fudenberg and Tirole 2010).

   a) **NIS (leader) Objective Function:** The NIS maximizes its profit by setting the price per site, with Capex in Year 0 and Opex annually from Year 1 to Year 9. Assuming NIS has the production capacity to supply as many sites as required. NIS anticipates the MNO's response when setting the price $M_s$ followed by OEM as follower and maximizing own profit for year 0.

$$\max \Pi_{sr} = \text{Sites}_t \times M_s + \sum_{t=1}^{N-1} \frac{\text{Sites}_t \times M_s \times \text{OpexRate}_{rt}}{(1+d)^t}$$

(17)

**Subject to:**

**Predatory Price constraint:** This constraint ensures that NIS does not set prices above the predatory threshold, preventing the MNO from switching to direct procurement from OEMs or adopting ORAN. The proxy to θ is the change in the Opex rate MNO needs to pay for ORAN.

$$M_s \leq M_{s,\text{predatory}} = M_o (1 - \theta)$$

(18)

Where:

- $M_o$: Price per site if the MNO were to procure directly from OEMs.

- θ: probability that the MNO switches to direct OEM procurement or adopts ORAN

b) **MNO's Objective Function:** The MNO's problem is to maximize its discounted profit stream, given the NIS's infrastructure deployment i.e. the number of sites to install based on the price $P_s$ set by the NIS as leader from the range of margins possible.

$$\Pi_{mr} = (\text{ARPU}_{rt} \times U_{rt}) - (\text{Sites}_{rt} \times P_S) + \sum_{t=1}^{N-1} \frac{(\text{ARPU}_{rt} \times U_{rt}) - (\text{Sites}_{rt} \times P_S \times \text{OpexRate})}{(1+d)^t}$$

(19)

Where:

- $\text{ARPU}_{rt}$: Average Revenue Per User in region r at time t.

- $U_{rt}$: Number of users in region r at time t.

- Sites$_{rt}$: Number of sites procured by the MNO in region r at time t.

- P$_s$: Price per site set by the NIS.

- OpexRate: Operational expenditure rate.

- d: Discount rate

4. **Simulation Scenario**

The simulation in this study uses data across U.S. counties to evaluate the strategic incentives in the game-theoretic framework under MRAN (Traditional, Predatory) and ORAN (DirectOEM) scenarios. Key input datasets include demographic and economic indicators sourced from the U.S. Census Bureau, which provides granular county-level information such as population density and business activity through the County Business Patterns (CBP) program. These demographic variables are integrated to represent regional demand potential and population-driven variations in the marginal cost structures of deployment. Median household income data from the 2020 census is incorporated to estimate average revenue per user (ARPU) as a function of local household conditions.

On the supply side, data on broadband availability, speeds, and usage patterns are sourced from the Federal Communications Commission (FCC), providing crucial insights into regional infrastructure capabilities. The FCC's Measuring Broadband America reports supply detailed performance metrics for mobile broadband, enabling the differentiation of network quality and demand elasticity across regions. Infrastructure deployment cost estimates are derived from the FCC's catalog, which outlines costs specific to different technology deployments (e.g., RAN upgrades) and regional geographies. These cost parameters are crucial for capturing the varying operational expenditures in densely populated urban centers compared to sparsely populated rural areas. This diverse array of data sources, spanning demographic, economic, and telecom-specific indicators, enables a multifaceted analysis of the potential for ORAN adoption across various region types in the United States (see Supplementary Material for data source and capacity calculations), illustrating how solutions vary across different geographies based on cost structures and regional characteristics for United States.

**County Clustering into Region Types**

First, U.S. Counties are categorized into eight distinct region types that recognize the inherent regional heterogeneity in terms of existing cell and population densities. Cell density serves as a proxy for the availability of telecommunications infrastructure per km in each county. We clustered all U.S. counties into eight (8) region types using k means clustering by using existing cell density and population density. These region types facilitate a more representative analysis by grouping them into distinct clusters. We employed the Elbow method and the Calinski-Harabasz Index to determine the optimal number of clusters, which is defined as three, including urban, suburban, and rural areas, indicating a clear point of diminishing returns in reducing within-cluster variance. However, the Calinski-Harabasz Index, which measures the ratio of between-cluster dispersion to within-cluster dispersion, continues to increase up to ten clusters, suggesting that more distinct groupings could be identified with a higher number of clusters. Given this discrepancy and considering our research objectives to capture a more nuanced spectrum of urban, suburban, and rural areas, we opted for eight clusters.

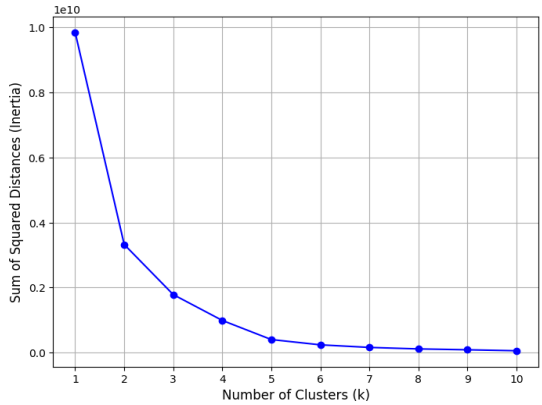

(a) Elbow Method Results

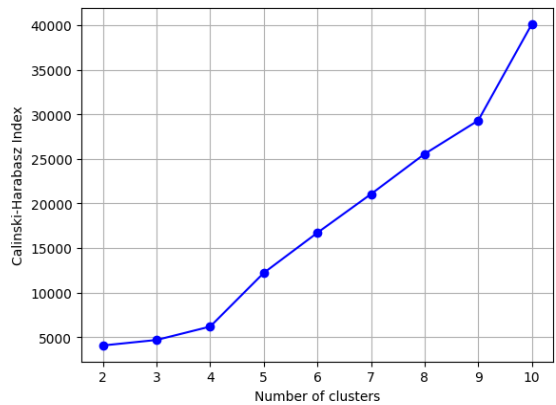

(b) Calinski-Harabasz Method Results

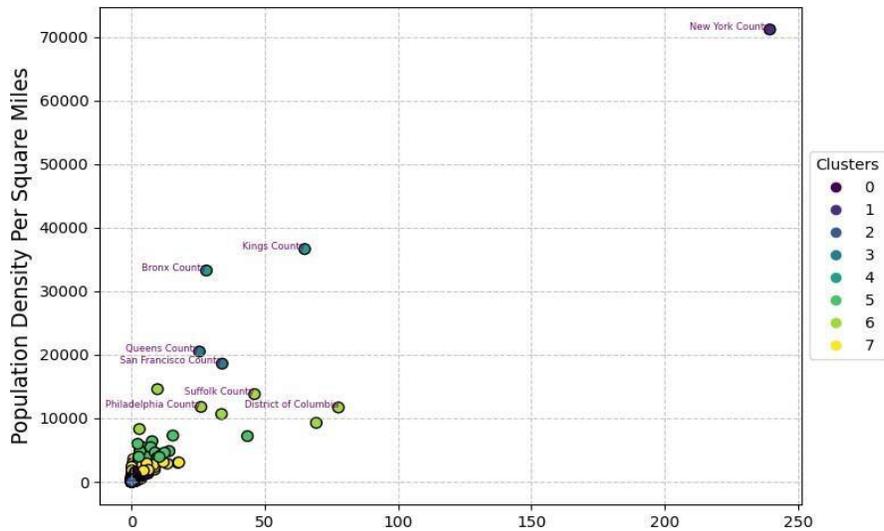

(c) K-Means Clustering for all Counties by Population Density and Cell Density

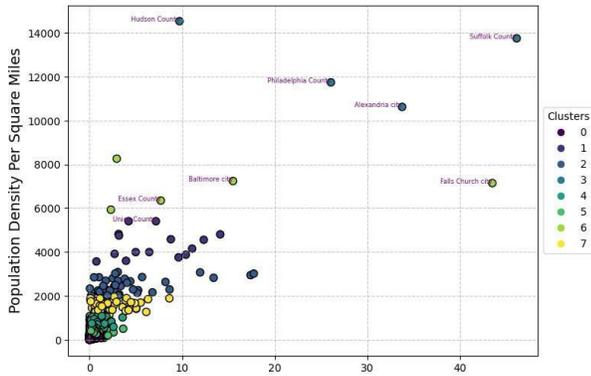
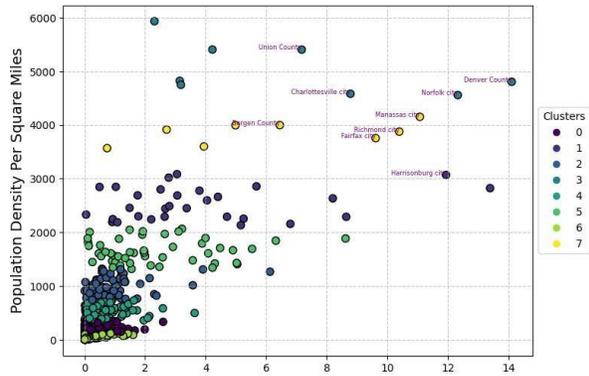

(d) K-Means Clustering (8SD, Z score)　　　　　　　(e) K-Means Clustering (7SD, Z score)

Figure 2- Comparison of Clustering Methods for K-means Clusters with Results

This diverse array of data sources, spanning demographic, economic, and telecom-specific indicators, enables a multifaceted analysis of the potential for Open RAN adoption across various region types in the United States.

| Category | Variables | Sources |
| --- | --- | --- |
| **Demographics & Household** | Population, Growth Rate | U.S. Census Bureau, 2020 Census Population (U.S. Census Bureau 2020a) |
| | Land Area | U.S. Census Bureau, 2023 Gazetteer Files (U.S. Census Bureau 2023) |
| | Businesses, Growth Rate | U.S. Census Bureau, County Business Patterns (CBP) Survey (U.S. Census Bureau 2020b) |
| | Median Household Income (MHI) | USDA ERS (U.S. Department of Agriculture Economic Research Service 2020) |
| **Network** | Existing Telecom Sites | Unwired labs OpenCellID (Unwiredlabs 2020) |
| | Data Demand & Speed Demand | FCC (Federal Communications Commission 2020) |
| **Cost** | FCC final catalog | FCC (Federal Communications Commission 2021) |

Table 2 - Datasets used for the analysis, including variables and sources.

The results from cluster analysis demonstrate distinct clusters among counties based on cell density and population density, with notable high-density outliers, such as New York County and Kings County, clearly labeled for reference. Also, the scatter plot Figure 2 visually presents these classifications, highlighting the differences in population density and existing telecom cell densities across regions. The urban areas show high density, while the gradual decline into suburban and rural regions illustrates the varying infrastructure needs. Also, it shows that population density is the real driver of infrastructure.

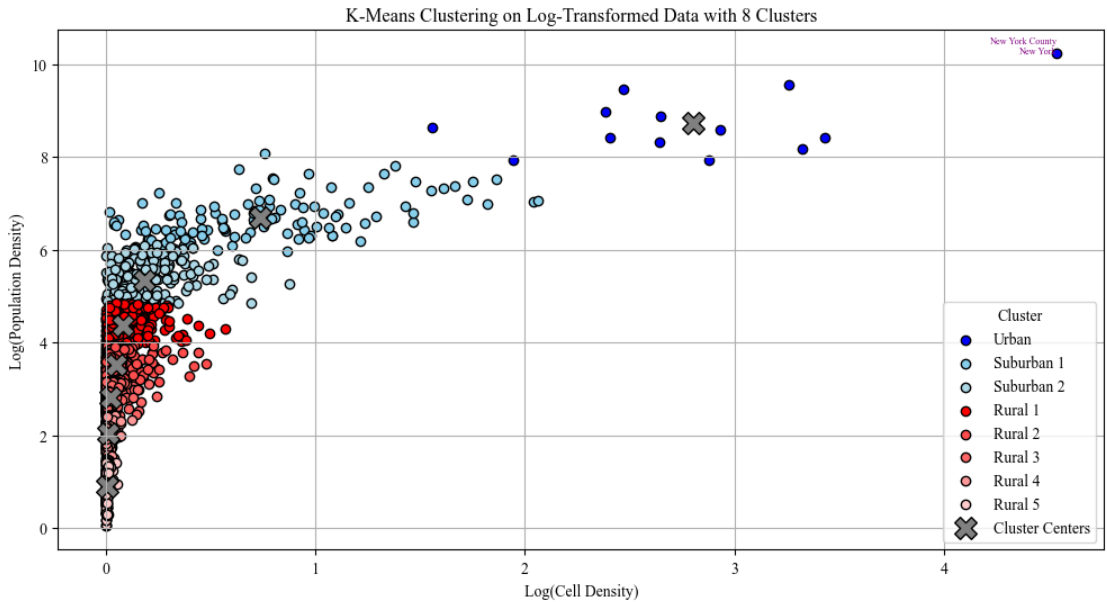

Figure 2 - Region Types Based on Population Thresholds

The resulting classification system comprises eight categories based on cell and population density, enabling us to account for varying infrastructure needs across different population densities (see Table 3 for detailed cluster centroids).

| Cluster | Cell Density Per sq km | Population Density Per sq km |
|---|---|---|
| **Urban** | 15.46 | 6185.54 |
| **Suburban 1** | 1.09 | 799.39 |
| **Suburban 2** | 0.20 | 208.56 |
| **Rural 1** | 0.09 | 77.73 |
| **Rural 2** | 0.05 | 32.89 |
| **Rural 3** | 0.03 | 15.66 |
| **Rural 4** | 0.01 | 6.77 |
| **Rural 5** | 0.01 | 1.45 |

Table 3 Cluster Centroids for Population Density Classification

Urban areas are categorized into Urban 1 (e.g., metropolitan cities like New York), with very high population and cell density, and Suburban 1 with high population and moderate cell density, both requiring advanced RAN technologies like massive MIMO. Suburban 2 represents transition zones with moderate population and cell density. The rural gradient includes Rural 1 (semi-rural edges with mixed use), Rural 2 (small towns with light agriculture), Rural 3 (low-density farming areas), Rural 4 (vast farmlands), and Rural 5 (remote wilderness), each reflecting decreasing population and cell densities. These classifications guide tailored telecom infrastructure strategies. This detailed categorization allows for a nuanced approach to understanding and addressing infrastructure needs from densely populated urban areas to sparsely populated rural zones. The significant variations in cell and population densities across these clusters highlight the diverse challenges in implementing adequate wireless network coverage across different regions.

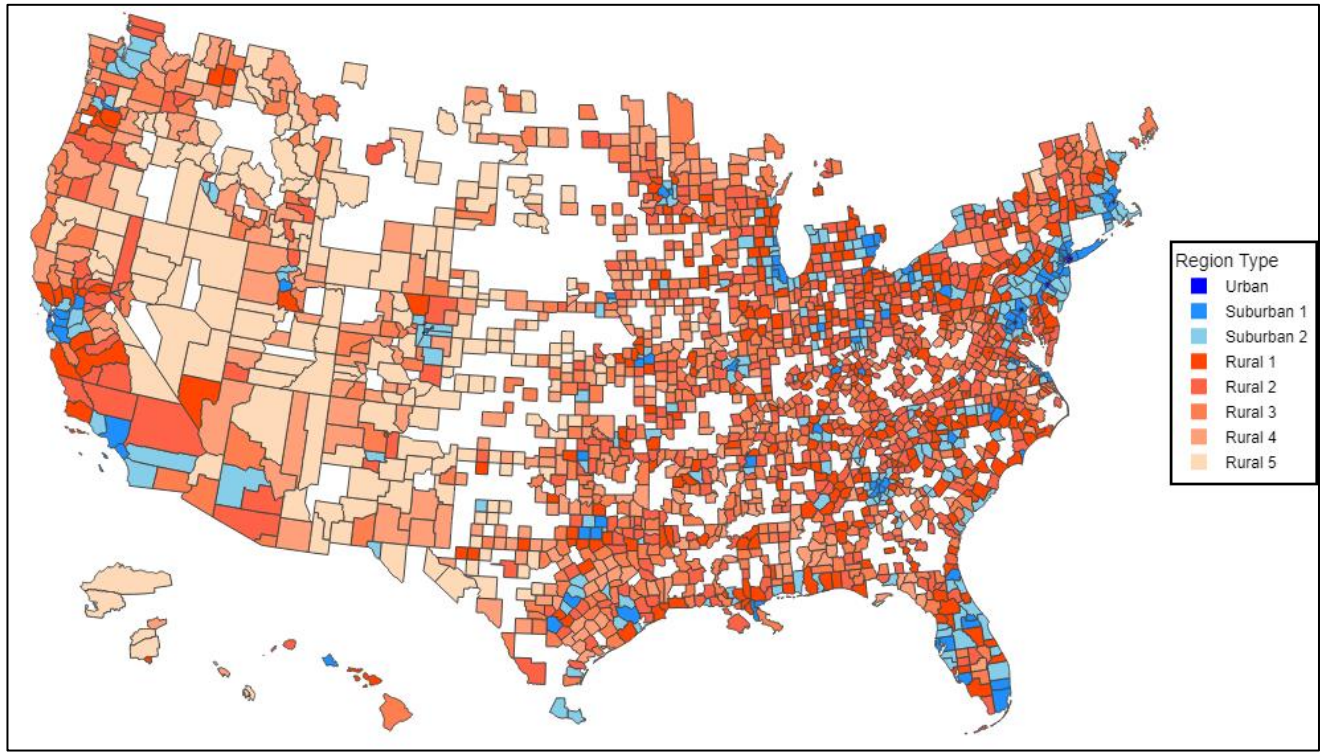

Figure 3 - Map of U.S. Counties Categorized by Region Type (by population and cell density)

**Configuration Selection and Capacity Specifications**

The configurations used in this study are selected to meet the distinct capacity and coverage requirements of Urban/Suburban1 and Suburban2, rural regions in USA counties. For urban and suburban areas, both the Traditional/Predatory and Direct OEM Pricing model provides MRAN and utilize high-capacity configurations setups i.e., the Traditional/Predatory models rely on multi-band LTE antennas with 10T10R to 20T20R configurations, enhanced by mMIMO eNodeB (64T64R) with 60 MHz bandwidth for advanced mobility features, ensuring high throughput in densely populated areas. The Direct OEM provides ORAN configuration and enhances this further with 5G Sub6 (100 MHz) carriers, providing low-latency, high-speed connectivity and supporting large user bases in urban settings.

In rural regions, the focus shifts toward coverage and cost efficiency. The Traditional/Predatory models use non-MIMO eNodeB with 20 MHz/sector and Fixed Wireless Access (FWA) features, ensuring wide area coverage with fewer sites. The Direct OEM model also employs a Macrocell with 4G 20 MHz carrier, providing sufficient capacity for low-density areas. Both configurations offer power flexibility (20W to 320W) to adapt to various terrain conditions. They are well-suited for the sparse population and large geographical areas typical of rural regions. These configurations balance cost, coverage, and capacity, ensuring optimal service across diverse county types.

**Configuration-based Capacity and Coverage Summary**

Table 4 summarizes the capacity and coverage that each configuration type provides. The selected configuration determines the bandwidth, spectral efficiency, and number of sectors, where higher bandwidths and spectral efficiencies allow for greater capacity. In urban/suburban regions, the use of mMIMO technology with configurations such as 64T64R improves capacity and spectral efficiency but can lead to a reduction in coverage area due to signal interference and obstacles in dense environments. This is why the coverage area in these regions remains at 7.07 sq km, despite high spectral efficiency. In contrast, rural areas rely on non-MIMO configurations, which cover larger areas (up to 176.71 sq km), although with lower spectral efficiency in traditional models.

Direct OEM configurations further enhance capacity, especially in urban regions, with 100 MHz bandwidth and 50 bps/Hz spectral efficiency. In comparison, rural regions benefit from improved spectral efficiency (20 bps/Hz) compared to traditional setups. However, rural configurations prioritize wide coverage over maximum data throughput, making them more cost-effective for low-density regions where fewer users need high-speed data access.

| Configuration, Region | B.W. (MHz) | Spectral $\eta$ (bps/Hz) | Sectors | Coverage (sq km) |
|---|---|---|---|---|
| MIMO, Urban/Suburban1 | 60 | 40 | 3 | 7.07 |
| non-MIMO, Rural | 20 | 10 | 1 | 176.71 |
| MIMO, Urban/Suburban1 | 100 | 50 | 3 | 7.07 |
| non-MIMO, Rural | 20 | 20 | 1 | 176.71 |

Table 4- Capacity and Coverage Specifications

**Assumptions and Key Parameters**

Due to the lack of data on invoice, bill of materials expenditure on infrastructure per county, the analysis relies on the FCC catalog, which provides the closest available estimates for costs, allowing for standardized cost assumptions across regions. We assume a 10-year investment horizon, as the typical lifespan of RAN infrastructure is 10 years, after which significant upgrades or replacements are required. All costs and revenues are projected over this period to reflect the infrastructure's operational lifecycle. Also, we apply a 5% discount rate to calculate the NPV of future cash flows. Also, for this study, inference is assumed to be 0 for all scenarios and region types.

In the Traditional scenario, MNOs engage with both NIS and OEMs for service delivery. The predatory scenario reflects a pricing strategy where MNOs bid for Open RAN, prompting network suppliers to offer predatory pricing to counter the competition and maintain market dominance. The DirectOEM scenario bypasses the role of network suppliers, enabling MNOs to directly procure equipment and services from OEMs.

MNO cost assumptions are derived from the FCC catalog Table 5. The base costs are assumed to differ by region type, with urban and Suburban 1 regions incurring significantly higher deployment costs compared to Suburban 2 and all Rural 1-5 areas. For the Traditional scenario, the initial setup cost of deploying telecom infrastructure is $232,823 per site in Urban and Suburban 1 regions as per the FCC MIMO catalog for MIMO technologies that suits the population densities, and $63,237 for Suburban 2 and Rural 1-5 regions. In contrast, for the predatory and DirectOEM scenarios i.e ORAN, MNO costs are significantly reduced to $108,000 and $42,500 per site for Urban/Suburban 1 and Suburban 2 and all Rural 1-5 regions, respectively, based on the ORAN configurations chosen.

Operating expenses (Opex) are assumed to be a fixed percentage of Capex, varying across scenarios. In the Traditional and predatory scenarios, Opex is set at 12%, whereas the DirectOEM scenario incurs slightly higher Opex, at 13%. These Opex rates account for the ongoing operational and maintenance costs associated with telecom infrastructure deployment and service provision.

The summary statistics for each region type in Table 5 provide insights into key variables such as population, MHI, data consumption, and existing infrastructure density. We have a balanced dataset with a total of N=2074 U.S. counties.

| Mean | Counties (N) | Population (k) | Median Household Income ($k) | Data (GB/month) | Cell Density per Sq Km (mean) | Pop. Density Per Sq km min-max(mean) |
|---|---|---|---|---|---|---|
| Urban | 13 | 762 | 85 | 168 | 21.191 | 2,760-27,469 (7,899) |
| Suburban 1 | 117 | 637 | 78 | 96 | 1.361 | 391-3,190 (893) |
| Suburban 2 | 240 | 258 | 76 | 36 | 0.215 | 126-445 (222) |
| Rural 1 | 334 | 91 | 65 | 10 | 0.091 | 51-128 (80) |
| Rural 2 | 466 | 44 | 59 | 6 | 0.052 | 23-50 (34) |
| Rural 3 | 441 | 23 | 55 | 3 | 0.025 | 10-23 (16) |
| Rural 4 | 279 | 15 | 55 | 2 | 0.014 | 3-10 (7) |
| Rural 5 | 184 | 8 | 57 | 1 | 0.005 | 0-3 (2) |

Note: k = thousands

Table 5- Summary Statistics by Region Type

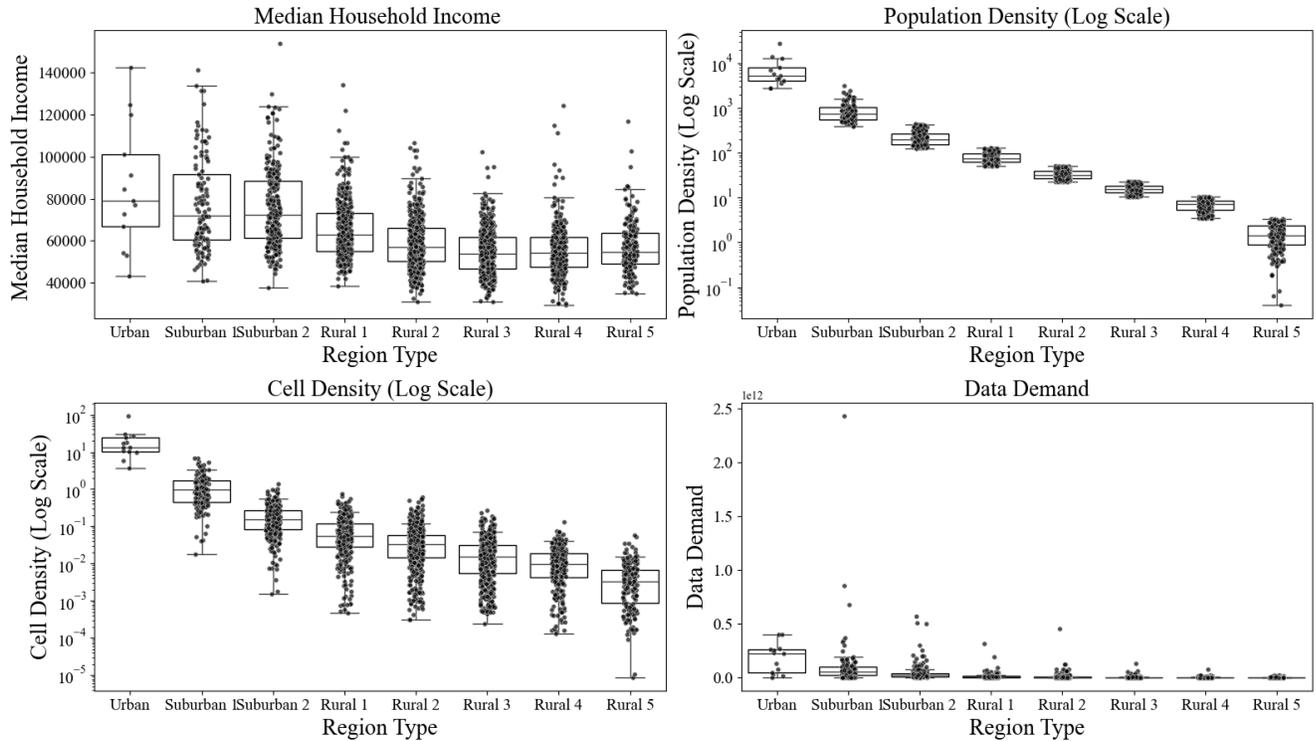

Figure 4 - Graphical descriptive summary of key variables.

The descriptive summaries of variables shown in Table 5 and Figure 4 demonstrate significant urban-rural disparities. The data shows that Urban areas generate high revenues from small land areas. In contrast, Rural regions produce lower revenues across larger territories, highlighting the varying market dynamics across the US. This pattern reflects broader economic disparities between urban and rural areas. The considerable variability within each region type also indicates that telecommunication companies should tailor their strategies to specific regional conditions rather than applying a one-size-fits-all approach. This provides a foundation for developing targeted Open RAN investment strategies that consider the diverse economic landscapes across different U.S. region types.

To estimate the revenue for MNOs in each region type, we consider two primary factors: the total number of active smartphone users and the ARPU specific to that region type. Our approach utilizes population data from the 2020 U.S. Census, providing accurate county-level demographics. For user adoption, we reference the GSMA North America 2020 report, which indicates an overall smartphone penetration rate of 83% for North America. ARPU data is derived from the CTIA annual wireless industry survey report, which finds an average ARPU of $35.74 for the United States in 2021.

To account for regional economic disparities, we weighed this national ARPU using the MHI of each county. This normalization process allows for a more accurate representation of revenue potential across diverse regions. This results in adjusted ARPU values for each of our defined region types: Urban, Suburban 1, Suburban 2, and Rural 1 through 5. These adjusted ARPU values reflect the varying economic conditions and potential revenue generation across different region types, providing a more nuanced basis for our revenue estimations in the Open RAN investment analysis. All other uncertainties and assumptions taken in the model are highlighted in section 4.3 Assumptions and Key Parameters and listed in supplementary Table 6 - List of Parameters Used in Model as the model inputs.

| General Parameters | | | |
|---|---|---|---|
| Parameter | Value | Description | Units |
| Investment Horizon | 10 | Typical for Telecom Capex | years |
| Discount Rate | 5 | Industry-Standard | % |
| Players | MNO, NIS, OEM | Key Players in RAN deployment | entities |
| Procurement Scenarios | Traditional, Predatory, DirectOEM | Procurement Scenarios under any RAN type | names |
| Smartphone User % | 89 | Smartphone users by GSMA NA 2020 | % |
| Active User % (average) | 83 | Active users on the network GSMA NA 2020 | % |
| ARPU (average) | 40 | Average Revenue Per User(ARPU), CSIA Survey 2020 | $ |
| Price Parameters | | | |
| Parameter Region | Traditional | Predatory | DirectOEM | Source |
| MNO Cost/NIS Price($) Urban/Sub | 232,823 | 108,000 | 108,000 | FCC |
| Rural | 63,237 | 42,500 | 42,500 | FCC |
| Opex Rate | 0.12 | 0.12 | 0.13 | TIP |

Table 6 - List of Parameters Used in Model

**Simulation Results**

This analysis delves into the transition from MRAN to ORAN, assessing its alignment with the study's objectives of improving cost efficiency, addressing infrastructure disparities, and enhancing profitability across a diverse set of regions. By overcoming the limitations of traditional MRAN models, such as higher costs and vendor lock-in, ORAN offers a scalable, flexible, and cost-effective solution. The results validate ORAN's potential to create a more competitive ecosystem, particularly in addressing the unique challenges posed by varying regional demand and infrastructure requirements.

The quantitative results highlight a consistent improvement in MNO payoffs across all region types when transitioning from MRAN to ORAN. In urban areas, the Net Present Value (NPV) of MNO payoffs increased by 10%, rising from $30 million under MRAN to $33 million under ORAN, driven by higher demand density and advanced network optimization. Similarly, in Suburban 1 regions, payoffs rose from $24 million to $26.4 million, also reflecting a 10% gain, emphasizing the impact of ORAN in regions with moderate infrastructure investments. Even in rural areas, which typically face significant cost and demand challenges, ORAN adoption delivered comparable efficiency gains. For instance, Rural 1 payoffs increased by 10%, from $12 million under MRAN to $13.2 million under ORAN, show casing ORAN's ability to bridge the profitability gap in low-demand regions.

These findings underscore ORAN's transformative role in modernizing telecommunications infrastructure. By reducing infrastructure costs and fostering scalability, ORAN enables MNOs to align regional investment strategies with profitability goals effectively. The results further highlight ORAN's ability to create a balanced and competitive landscape, benefiting not only high-demand urban centers but also underserved rural areas, making it a critical driver for equitable and efficient network deployment.

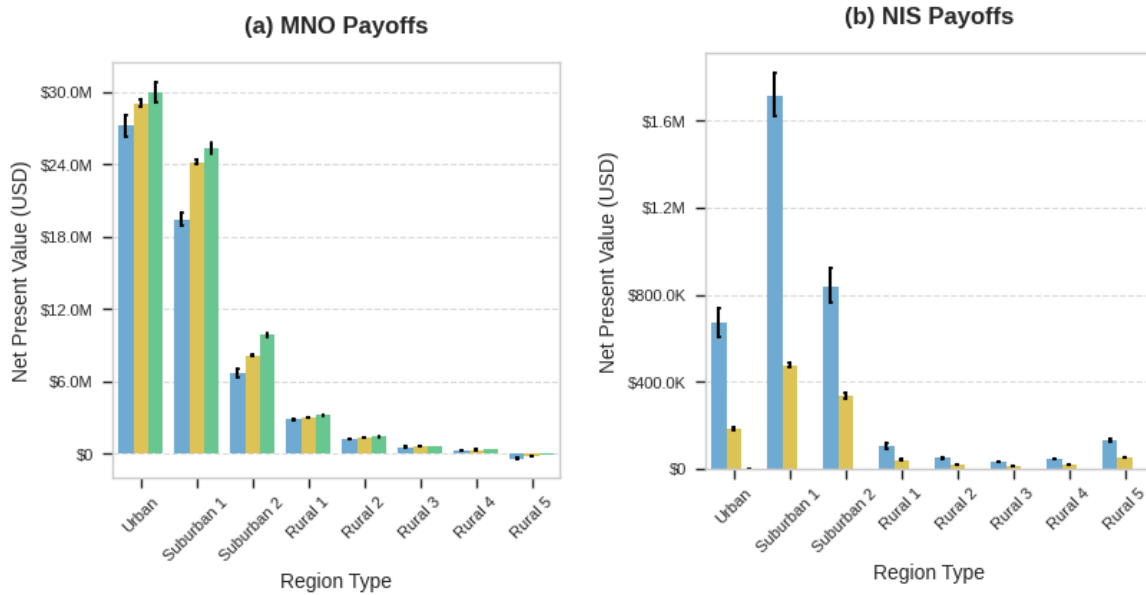

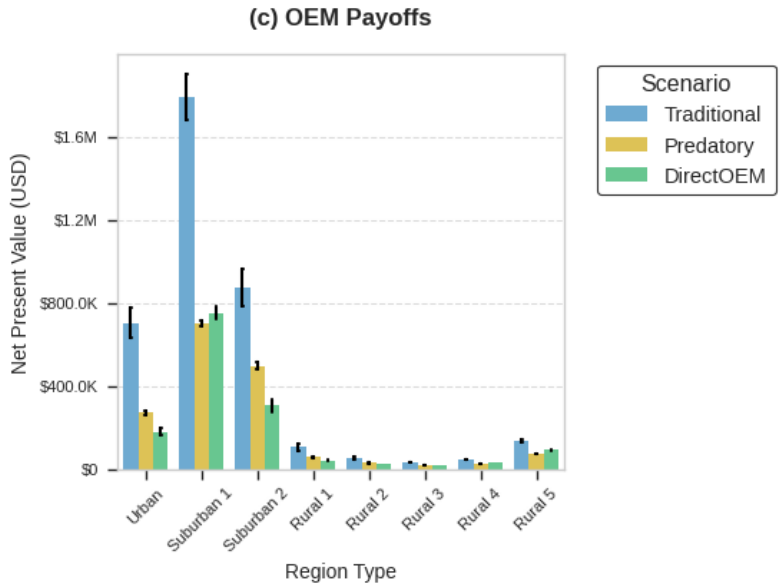

Figure 5 - Comparison of MRAN (Traditional and Predatory) with ORAN payoffs as Net Profit

Identifying a global equilibrium involves finding optimal margins for different stakeholders (e.g., MNOs, NIS, and OEM) such that total profits are maximized across all scenarios. We employed a Random Forest Regressor to identify the optimal margins for NIS and OEM that maximize the total profit across different scenarios and regions. The input features include the scenario, region type, year, NIS margin, and OEM margin. The Random Forest model was trained using 100 decision trees. After training, a Gaussian process optimization to identify the NIS and OEM margin values that maximized predicted profits. The objective function minimized the negative predicted profit, effectively searching for the margins that would yield the highest cumulative profit across the different stakeholders. The results from the Random Forest model optimization highlight significant insights into profit distribution across stakeholders in the telecom supply chain.

The optimization process, conducted over 50 iterations, identified the optimal NIS and OEM margins. The analysis determined the optimal NIS margin to be approximately 41.86% and the OEM margin to be around 17.34%. These optimal margins correspond to the highest predicted total profit across all stakeholders, suggesting important strategic implications. This model provides a robust method for determining optimal pricing strategies across multiple deployment scenarios and regions, contributing to more informed decision-making in telecom network planning and investment with key insights.

First, the lower margin for OEMs reflects increasing cost pressures in equipment manufacturing as the market shifts towards more decentralized and open network systems like ORAN. This may prompt OEMs to adjust their pricing models or seek alternative revenue streams to maintain profitability amidst evolving industry dynamics. Finally, these insights reinforce the importance of strategic margin balancing in contract negotiations and partnerships. Telecom firms can enhance profitability by securing cost-effective OEM contracts while allowing network integrators to maintain higher margins. This approach aligns with the emerging trend of decentralization and could drive better financial performance in both traditional and DirectOEM deployment scenarios.

While our model focuses primarily on deployment and operational costs, it is important to note that security considerations in ORAN implementations can impact both performance and costs. Recent research has shown that encryption on critical ORAN interfaces can add latency and limit throughput. While not explicitly factored into our current cost model, these security-related performance impacts should be considered in future, more detailed analyses of ORAN deployments (Groen, Kim, and Chowdhury 2023).

**Supplementary Material**

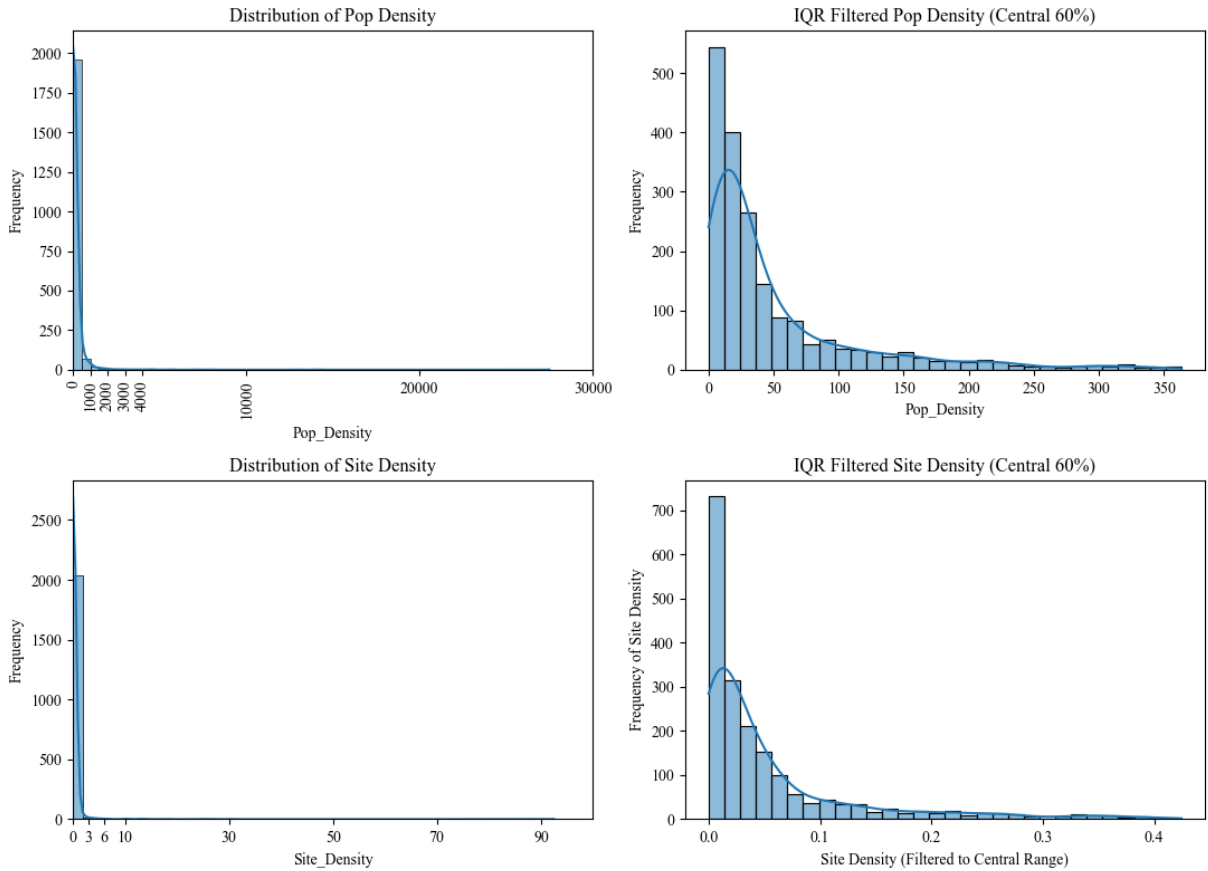

Figure 6 - Distribution of Density Plots

| FCC Catalog - MRAN Tech Spec. - Traditional Mono RAN | | | |
|---|---|---|---|
| Description | Low ($) | High ($) | Average ($)* |
| Antenna - LTE (Long-Term Evolution) Multi-band, >16dBi, 10 port - 10T10R through 20 port - 20T20R | 1,479 | 10,995 | 6,237 |
| mMIMO eNodeB 3 Sector Per Band (64T64R 20 MHz 1 Band FWA - 64T64R 60 MHz 1 Band with Advanced Mobility Fea- tures) | 85,000 | 368,172 | 226,586 |
| **Total for Urban and Suburban** | **86,479** | **379,167** | **232,823** |
| Antenna - LTE (Long-Term Evolution) Multi-band, >16dBi, 10 port - 10T10R through 20 port - 20T20R | 1,479 | 10,995 | 6,237 |
| non-MIMO eNodeB with 3 sectors, single spectrum band of up to 20 MHz/sector. Fixed Wireless features. Range due to Low Power radio (20W) vs High Power radio (up to 320W). Price includes RRHs, BBU, Ancillaries, Software Features, and Ca- pacity Licensing. Price excludes antennas, tower cabling, tower ancillaries and over voltage protection (OVP) | 45,000 | 69,000 | 57,000 |
| **Total for Rural** | **46,479** | **79,995** | **63,237** |

| **FCC Catalog - ORAN Tech Spec.** | | | |
|---|---|---|---|
| **Description** | **Low ($)** | **High ($)** | **Average ($)*** |
| Macrocell - 5G Sub6 100MHz Carrier (Bundle - Antennas, RU/RRU, BBU Software, GPS Receiver, STU, RIU, Mechani- cal Mounting, Cables and Connectors, SFPs) | 91,000 | 125,000 | 108,000 |
| **Total for Urban and Suburban** | **91,000** | **125,000** | **108,000** |
| Macrocell 4G 20MHz Carrier (Bundle - Antennas, RU/RRU, BBU Software, GPS Receiver, STU (Subscriber Terminal Unit), RIU (Radio Interface Unit), Mechanical Mounting, Cables and Connectors, SFPs) | 39,000 | 52,000 | 45,500 |
| **Total for Rural** | **39,000** | **52,000** | **45,500** |
| *Averages are used in the simulation model* | | | |

Figure 7- FCC Catalog Comparison Tech Spec and Price- Traditional Mono RAN (MRAN) vs. ORAN (ORAN) RAN Site

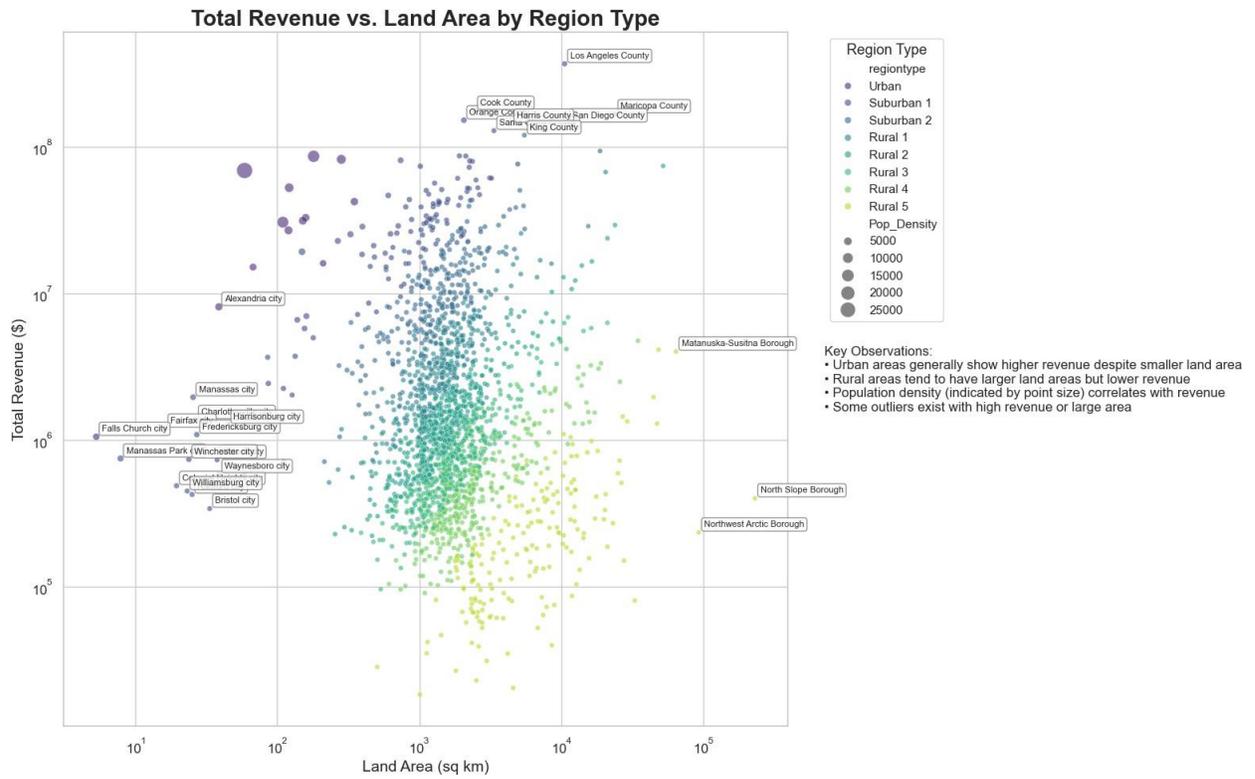

Figure 8 - Total Revenue vs Total Land Area per County